\documentclass[aps,prd,twocolumn,showpacs]{revtex4}
\usepackage{amsmath}
\usepackage{epsfig}
\usepackage{amsfonts}
\begin{document}
\def\b{\bar}
\def\d{\partial}
\def\D{\Delta}
\def\cD{{\cal D}}
\def\cK{{\cal K}}
\def\f{\varphi}
\def\g{\gamma}
\def\G{\Gamma}
\def\l{\lambda}
\def\L{\Lambda}
\def\M{{\Cal M}}
\def\m{\mu}
\def\n{\nu}
\def\p{\psi}
\def\q{\b q}
\def\r{\rho}
\def\t{\tau}
\def\x{\phi}
\def\X{\~\xi}
\def\~{\widetilde}
\def\h{\eta}
\def\bZ{\bar Z}
\def\cY{\bar Y}
\def\bY3{\bar Y_{,3}}
\def\Y3{Y_{,3}}
\def\z{\zeta}
\def\Z{{\b\zeta}}
\def\Y{{\bar Y}}
\def\cZ{{\bar Z}}
\def\`{\dot}
\def\be{\begin{equation}}
\def\ee{\end{equation}}
\def\bea{\begin{eqnarray}}
\def\eea{\end{eqnarray}}
\def\half{\frac{1}{2}}
\def\fn{\footnote}
\def\bh{black hole \ }
\def\cL{{\cal L}}
\def\cH{{\cal H}}
\def\cF{{\cal F}}
\def\cP{{\cal P}}
\def\cM{{\cal M}}
\def\ik{ik}
\def\mn{{\mu\nu}}
\def\a{\alpha}

\title{Source of the Kerr-Newman Solution as a Supersymmetric Domain-Wall Bubble:\\
50 years of the problem}
\author{Alexander Burinskii}

\affiliation{Theor. Phys. Lab.  NSI, Russian Academy of Sciences,
B. Tulskaya 52  Moscow 115191, Russia\footnote{E-mail address:
bur@ibrae.ac.ru}}

\begin{abstract}
We consider the chiral field model of the source of the Kerr-Newman (KN) solution and obtain that
it represents a supersymmetric spinning soliton, bounded by the chiral domain wall (DW) of the
ellipsoidal form. The known method for transformation of the planar DW to Bogomolnyi form we
generalize to the curved DW-bubble adapted to the Kerr coordinate system and obtain the
supersymmetric BPS-saturated source of the KN solution, having some remarkable features, in
particular, the quantum angular momentum. The main new result is that the source forms a breather,
i.e. the DW-antiDW combination. Taking into account that the KN solution describes the spinning
particles with gyromagnetic ratio $g=2 ,$ as that of the Dirac electron, we touch  the problem of
the compatibility of the spinning particles with gravity.
  \end{abstract}


\maketitle

Last year we marked 50 years of the problem of source of the
Kerr-Newman (KN) solution. Starting in 1965 in the paper by Newman
and Janis  \cite{NewJan}, where it was mentioned that the KN space
is two-sheeted, the diverse attempts and  suggestions to solve
this problem are debating up to now \cite{Hehl,Dym,BurBag}. In
this paper we develop the line started by W. Israel \cite{Isr},
who suggested to truncate the second sheet of the Kerr geometry
along the \emph{disk} spanned by the Kerr singular ring. After
analysis of the Israel source by Hamity \cite{Ham}, a modified
\emph{disk-like source} was suggested by C. L\'opez \cite{Lop} as
an ellipsoidal vacuum bubble -- a thin shell covering the Kerr
singular ring and matching with the external KN solution. In the
papers \cite{BurBag,BurBag2} we considered a generalization of the L\'opez model, in which
the  thin shell is replaced by a \emph{domain-wall} (DW) forming a
spinning bubble-soliton \cite{BurSol}, based on the N=1
supersymmetric chiral field model.\fn{Earlier story of the problem
of the KN source was described by W.Israel \cite{Isr}, and later
by A.Krasinski \cite{Kras}.} The corresponding
Hamiltonian was reduced to Bogomolnyi form, and it was shown that this soliton forms a
supersymmetric, PBS saturated state.  In this letter we would like
pay attention to the used intricate method reduction to Bogomolnyi form,
which was apparently first suggested in \cite{FMVW} for a two-dimensional kink solution,
and then successfully used for the planar DW in
\cite{AbrTown,CvQRey,GibTown,Shif,ChibShif}.
We generalize this method to the much more complex case of
 the Kerr geometry, in which the source is spinning and bounded by the DW
 of  ellipsoidal form. Besides, it is formed by the system of chiral
 fields, when one of them depends on the Kerr angular coordinate and time.
With respect to previous treatment \cite{BurBag,BurBag2}, we obtain  new very
important feature of the DW source -- formation  of the DW-antiDW (breather) structure.

In addition to the purely academic interest, the problem of the source
of the KN solution is also important for resolution of the
conflict between quantum theory and gravity. As is known, the KN
solution has the gyromagnetic ratio of the Dirac electron
\cite{Car}, and therefore, it corresponds to the external
gravitational and em field of the electron \cite{DKS}. Structure
of the source of the KN solution could shed the light on the
question, which peculiarities of the spinning particles provide
their consistency with gravity, and we come to the conclusion that
the supersymmetric DW phase transition  is one of the conditions
for this consistency.

\textbf{Spinning soliton.}
\noindent Metric of the KN solution in the Kerr-Schild (KS) form
\cite{DKS}
 is \be g_\mn =\eta_\mn + 2H k_\m k_\n , \label{gKS}\ee
where $ \eta_\mn $ is metric of the auxiliary Minkowski
space\fn{We use signature $(- + + +)$ and the dimensionless units $G=c=\hbar=1 $.}  $\mathbb{M}^4 ,$  \be
H=\frac {mr -e^2/2}{r^2+a^2 \cos ^2 \theta}\label{HKN} \ee
 is a scalar function, expressed in the oblate ellipsoidal coordinates  $r$ and
$\theta ,$  related with the Cartesian coordinates $(t,x,y,z) \in
\mathbb{M}^4$ as follows \cite{DKS}
 \bea x +iy &=&(r +ia) e^{i\phi_K} \sin \theta, \nonumber \\
 z&=& r\cos \theta ,  \nonumber \\
 t &=&\rho - r  \label{Kerr coord}. \eea

 The null vector field
\be k_\m dx^\m = dr - dt - a \sin ^2 \theta d\phi_K,  \label{km} \ee
($ k_\m k^\m =0 $) is tangent to
  Principal Null Congruence (PNC) $\cal K $ of the Kerr geometry,
 forming a vortex-like polarization of the Kerr space-time.
Similarly to the metric, the vector potential \be A_\m dx^\m = -
Re \ [(\frac e {r+ia \cos \theta})] k_\m \label{Am} \ee is also
aligned to the PNC direction  $ k_\m .$

\begin{figure}[ht]
\centerline{\epsfig{figure=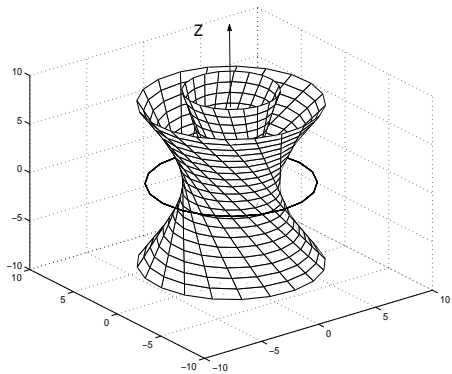,height=5cm,width=6cm}}
\caption{The Kerr congruence, generated by null vector field $k^\m ,$  is extended analytically through the
Kerr singular ring to `negative' sheet of the Kerr space, $r<0$,
creating two-sheeted space-time.}
\end{figure}

 The surface $r=0$ represents a disk-like
"door" from negative sheet $r<0$ to positive one $r>0 ,$ see
Fig.1. The null vector fields $k^{\m\pm}(x)$ turns out to be
different on these sheets, and two different null congruences
${\cal K}^\pm ,$ create two different metrics $ g_\mn^\pm
=\eta_\mn + 2H k_\m^\pm k_\n^\pm $ and two different EM field on
the same Minkowski background.

  The mysterious two-sheeted structure of the Kerr geometry created
   the problem of a ``realistic'' source of the KN solution.
  Relevant "regularization" of the KN solution was suggested by L\'opez \cite{Lop},
  who excised singular
  region together with negative sheet and replace it by a regular core
  with a flat internal metric $\eta_\mn .$
  The external KN metric (\ref{gKS})  matches with flat metric inside the bubble
  at the condition $H(r)=0 ,$ which determines
   the bubble boundary $r=R$ as the surface of ``zero-gravity
   potential'',
  \be
H|_{r=R}(r)=0 \label{HR0} .\ee

In accordance with (\ref{HKN}) it yields  \be R=r_e = \frac
{e^2}{2m} . \label{Hre} \ee
 Since $r$ is the Kerr oblate radial coordinate,  the bubble surface forms the disk with radius
 $r_c \sim a$ and thickness $r_e =e^2/2m .$ The matter sources and currents are distributed
over the surface of this disk, and the EM vector potential takes
maximal value in the equatorial plane $\cos
 \theta=0 ,$ at the disk boundary $r= r_e ,$
where the vector potential is dragged by the Kerr singular ring, see Fig.2, forming a closed loop
along the sharp border of the source.

\begin{figure}[h]
\centerline{\epsfig{figure=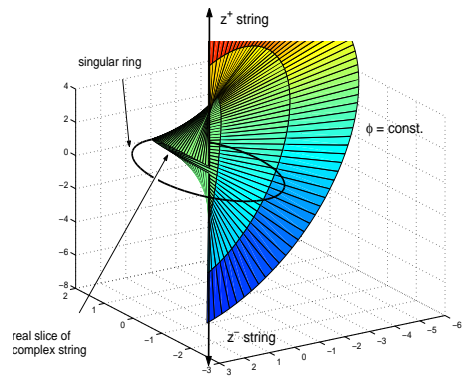,height=5cm,width=6cm}} \caption{\label{label} The Kerr
coordinate surface $\phi_K=const.$ Generators of the Kerr congruence $k_\m$ are tangent to the Kerr singular ring at $\theta=\pi /2 ,$
and vector potential forms a closed loop along the border of the KN source.}
\end{figure}

In \cite{BurSol} this model was extended  to a spinning
soliton-bubble model representing the domain-wall (DW)  interpolating between internal flat
space-time and the exact KN solution outside the bubble. It was assumed
\cite{BurSol,BEHM} that the regulated smooth KN solution is
described by a "deformed" KS metric suggested by G\"urses and
G\"ursey
 \cite{GG}, which retains the KS form (\ref{gKS}), but has the "deformed"  function

 \be H =
  H_{GG} =
 \frac{f(r)}{r^2 +a^2 \cos\theta}, \ee
 where $f(r)$ interpolates between zero value inside: $f=0$ by $r<r_e ,$
 and the exact KN form outside the bubble: \be f \equiv f_{KN}= mr -e^2/2
 .\ee

For cosmic black holes (BH) of large mass,  this source may be considered as interior of the BH
hidden by the BH horizon. For the KN solutions with  parameters of the elementary particles, the
spin/mass ratio is extremely large, $a>>m ,$ the horizons disappear, and the rotating source is
alternative to the naked singularity of the ultra-extreme KN solution \cite{BEHM,BurSol,BurBag}.

\textbf{\textbf{Supersymmetric phase transition.}}
As in many other soliton models, the considered in \cite{BurSol} soliton was formed by the Higgs mechanism of symmetry breaking. However, it was shown that the typical quartic potential term $V=g(\bar\sigma
\sigma - \eta^2)^2 ,$ where $\sigma$ is v.e.v. of
the Higgs field $\Phi(x)$, does not suit for the source of the
Kerr-Newman solution, since it leads to the location of the Higgs condensate outside the source that breaks the gauge symmetry of the external EM field. The correct scheme of the phase transition leading to location of
the Higgs field inside the source was used in \cite{BurSol}. It is  N=1 D=4 supersymmetric field model
\cite{WesBag}  with  three chiral fields $\Phi^{(i)}, \ i=1,2,3 ,$ forming a domain wall (DW) bubble.

The Lagrangian corresponding to bosonic part of the N=1
supersymmetric field model has the form \cite{WesBag} \be {\cal L}= -\frac
14 F_\mn F^\mn - \frac 12 \sum_i(\cD^{(i)}_\m \Phi^{(i)})(\cD^{(i)
\m} \Phi^{(i)})^* - V \label{L3} .\ee

Due to condition (\ref{HR0}) the DW boundary of the source may be considered on the flat
background, $g_\mn = \eta_\mn ,$ and the covariant derivatives
$\cD^{(i)}_\m \equiv \d_\m + ie A^i_\m $ are taken to be flat.

Among the three chiral fields $\Phi^{(i)}, \ i=1,2,3,$ we select the field $\Phi^{(1)}$ as the Higgs field $\Phi \equiv \Phi^{(1)} $,
and consider only one gauge field $A^1_\m \equiv A_\m  $ as the vector potential of the
KN EM field, setting $A^2_\m =A^3_\m =0 .$

The potential $V$ of the supersymmetric chiral model  is determined by
the superpotential \be V(r)=  \sum _i |\d_i W|^2 , \quad  \d_i W \equiv \d W /\d\Phi^i ,\label{VdW}\ee
 which must provide the necessary
concentration of the Higgs field inside the bag \cite{BurSol}. The corresponding superpotential was suggested by Morris in \cite{Mor} in the form  $W(\Phi^i, \bar \Phi^i) = Z(\Sigma
\bar \Sigma -\eta^2) + (Z+ \m) \Phi \bar \Phi,$ where
$ \m$ and  $ \eta $ are real constants, and
$ (\Phi, Z,
\Sigma) \equiv (\Phi^1, \Phi^2, \Phi^3) . $ The vacuum states
are determined  by the conditions  $ \d_i W =0 , $ which
yield:

\textbf{(I)} internal vacuum: $r<R-\delta $, $ V (r) = 0 , \
|\Phi| = \eta = const., \ Z=-\m, \ \Sigma =0, \ W_{in}=\m \eta^2 ,$ and

\textbf{(II)} external vacuum: $r>R$, $ V (r) = 0 , \
\Phi =0, \ Z=0, \ \Sigma=\eta, \ W_{ext} =0. $

We select also the phase transition zone $R-\delta < r < R $,

\textbf{(III)}  where the vacua
\textbf{(I)} and \textbf{(II)} are separated by a positive spike
of the potential $V ,$ and the surface currents are concentrated.

The part of the Lagrangian which is related with the nonzero Higgs field $\Phi \equiv \Phi^{(1)}$
is just the same as the Lagrangian used by Nielsen-Olesen in the model of the vortex string in
superconducting media
 \cite{NO}.  Similarly, the KN source may be considered as a
 disk formed of a superconducting matter.
 This part of the Lagrangian leads to the
 equations, \cite{BurSol,BurBag},
 \be \d_\n \d^\n A_\m = I_\m = e |\Phi|^2 (\chi,_\m + e
A_\m) \label{Main} \ee
 which describe  interaction between the complex Higgs field $\Phi(x) = |\Phi (x)| e^{i\chi(x)} $
and the KN  vector potential $A_\m $ penetrating inside the source. Regularization of the KN
electromagnetic field inside the source occurs via compensation of the vector potential $A_\m$ by
gradient of the phase of the Higgs field $\chi,_\m ,$ leading to the vanishing of the current
$I_\m$ inside the source and its localization on the border of DW, which is typical for
superconductors. In agreement with (\ref{Main}), the surface currents  compensate the difference
between the internal and external values of the vector-potential. As was shown in
\cite{BurSol,BurBag}, this mechanism leads to  two  important features of the source:

\textbf{(A)}: the Higgs field oscillates with
frequency $\omega= 2m ,$

\textbf{(B)}: angular momentum  is quantum, $ J=n/2,  n=1,2,3.$

\textbf{Bogomolnyi equations.}
Stress-energy tensor of the considered system consists of the pure
em part $T^{(em)}_\mn$ and contributions from the chiral fields
$T^{(ch)}_\mn$, \bea \nonumber T^{(tot)}_\mn = T^{(em)}_\mn +
\sum_i(\cD^{(i)}_\m \Phi^i)\overline {(\cD^{(i)}_\n \Phi^i)} \\
- \frac 12 g_\mn[\sum_i(\cD^{(i)}_\lambda \Phi^i)\overline
{(\cD^{(i)\lambda} \Phi^i)} +V] . \label{T3}  \eea
Flatness of the metric inside the bubble and in the vicinity of the domain
wall boundary leads to vanishing of the cross-terms in the metric
tensor, and using (\ref{VdW}) we can simplify the chiral part of
the Hamiltonian to the form \be H^{(ch)} \equiv \m = \frac
12 \sum_{i=1}^3 ~ [ \sum_{\m=0}^3 |\cD^{(i)}_\m \Phi^i|^2 + |\d_i
W|^2 ]. \label{HamCh} \ee
which represents the mass density of the domain wall $\m .$

Absence of the cross-terms indicates that the flat interior of the bubble and its DW boundary are not rotating.
However, the strong influence of gravity is saved in the
ellipsoidal shape of the DW and in the strong drag effect acting
of the KN electromagnetic field, which is to be aligned with twisted null direction
$k_\m$ of the Kerr congruence even in the limit of the flat space-time, see Fig.2.
Taking it into account, we use the  Kerr
coordinate system (\ref{Kerr coord}) which is adapted to the Kerr congruence  (\ref{km}) and to
ellipsoidal shape of the bubble-source.

In previous treatments \cite{BurSol,BurBag} we obtained that the Higgs field confined inside the
source
 depends on angular coordinate and time
\be \Phi \equiv \Phi^1(x) = |\Phi^1
(r)| e^{i\chi(t,\phi)} , \label{Phichi}\ee
while the other chiral fields depend only on $r .$
We separate in  (\ref{HamCh}) the terms depending on $\phi$ and $t,$
 \be  2 H^{ (ch)} = |\cD^{(1)}_t \Phi^1|^2 +
|\cD^{(1)}_\phi \Phi^1|^2 + \sum_{i=1}^3 ~ [ |\cD^{(i)}_r \Phi^i|^2 + |\d_i
W|^2 ], \label{HamChSep} \ee and note that they are positive definite, and we obtain two Bogomolnyi
equations which are valid for interior of the source,
 \be \cD^{(1)}_t \Phi^1
=0, \quad \cD^{(1)}_\phi \Phi^1= 0. \label{mainAH} \ee
One sees that for interior of the source, where $|\Phi^1|$ is constant and $J_\m=0 ,$ (\ref{mainAH}) is analog (\ref{Main}),
 \be \cD^{(1)}_\m \Phi^1 = i\Phi ( \chi,_\m + e A_\m)= (i/e\Phi)J_\m. \label{JD} \ee

Reduction of the last term is cumbersome and was  considered earlier for the planar chiral DW models in \cite{FMVW,AbrTown,CvQRey,GibTown,Shif,ChibShif}. There are introduced some constant phase factors
$e^{i\alpha_i},$  which allowed to rewrite  last term in the form
\bea \nonumber H^{(ch-r)} = \sum_{i=1}^3 \frac 12 | \cD^{(i)}_r \Phi^i - e^{i\alpha_i}\d \bar W /\d
\bar \Phi^i |^2  \\
+ Re \ e^{-i\alpha_i}(\d \bar W /\d \bar \Phi^i) \cD^{(i)}_r \Phi^i ,
\label{HamCh2} .\eea
These phases regarded as constants, which  looked somewhat artificial.
In the case of Kerr DW the phase $\alpha_1 $ turns out to be related with phase
of the Higgs field
and takes the physical meaning of the compensating field
for $A_\phi $ and $A_t $. Therefore, we allow this phase to be functions
of coordinates, except the coordinate $r $ that parametrizes  profile DW.
  From  (\ref{HamCh2}) we obtain  Bogomolnyi equation
 \be
  \cD^{(i)}_r \Phi^i - e^{i\alpha_i}\d \bar W /\d \bar
\Phi^i =0 , \label{Bog}\ee in which the coordinate $r$ parametrizes the  oblate  surface of the bag
similar to the parallel  surfaces of the planar domain wall, see Fig.3.

\begin{figure}[ht]
\centerline{\epsfig{figure=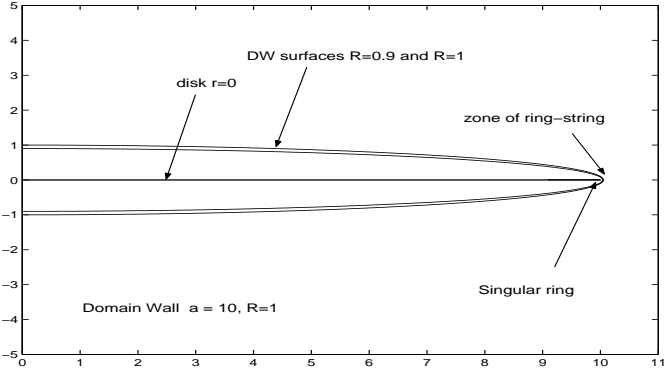,height=5cm,width=9cm}}
 \caption{Profile of the spheroidal domain wall phase transition.}
\end{figure}

The function $ W $ and $ Z $ are real, and without loss of generality, we can also set a real $\Phi^3 ,$ which allows us to take $\alpha_2 =\alpha_3 =0 .$  From  (\ref{Bog}) and (\ref{Phichi}) we obtain
$\alpha_1 = 2\chi (t,\phi) ,$  leading to $ \cD^{(1)}_r \Phi = e^{2i \chi (t,\phi)}  \cD^{(i)}_r \bar\Phi $ and
\be H^{(ch-r)} = \sum_{i=1}^3 \frac 12 | \d_r \Phi^i - \d W /\d
 \Phi^i |^2 +  Re \ (\d  W /\d  \Phi^i) \d_r \Phi^i ,
\label{HamChd} \ee where the replacement of the covariant derivatives $\cD^{(1)}_r$ to partial
$\d_r$ is valid due to concrete form of the used superpotential.

Minimum of the energy density $H^{(ch-r)}$ is achieved for
 \be \cD^{(i)}_r \Phi^i = \d  W /\d \Phi^i \quad \cD^{(i)}_r \bar\Phi^i = \d \bar W /\d \bar
\Phi^i , \label{Bog1}\ee which corresponds to saturated Bogomolnyi bound.
The expression (\ref{HamChd})  turns into full differential
\be H^{(ch-r)} =  Re \ (\d  W /\d  \Phi^i) \d_r \Phi^i =  \d W /\d r   . \label{HamChd1} \ee
The mass-energy of  the  supersymmetric interior of the bubble together with DW boundary
is
\be  M_{ch}= \int dx^3 \sqrt{-g} \ T_0^{\ 0
(ch)} , \ee where $\sqrt{-g}= (r^2 +a^2 \cos^2 \theta) \sin \theta.$
 Axial symmetry allows us to integrate over $\phi ,$ leading to
\be M_{ch}= 2\pi \int dr d\theta  (r^2 +a^2 \cos^2 \theta) \sin \theta \ T_0^{\ 0 (ch)} .
\label{MbubKerr}\ee
Using (\ref{HamChd1})  we obtain
\be M_{ch}= 2\pi \int dr d\theta  (r^2 +a^2 \cos^2 \theta) \sin \theta  \d_r W    .
\label{MbagKerr}\ee
Taking into account that superpotential  $W (r)$ is constant inside and outside the source,
 \be W_{int} = \m\eta^2, \ W_{ext}=0, \ee
we have $\d_r W =0 $ inside and outside the bag and, by crossing the bag boundary, we get the incursion
$\Delta W = W(R+\delta) -  W(R -\delta) =  -\m \eta^2 $ .
After integration over $r\in [0,R] ,$ and then over $X=\cos\theta$
we obtain
\be M_{ch}= 2\pi \Delta W \int_{-1}^{1}  dX  (R^2 +a^2 X^2 ) = 4\pi(R^2  +\frac 13 a^2) \Delta W .\label{MbagKN}\ee

However, working in the very specific Kerr coordinate system, we note that coordinates extend
analytically to negative $r,$ enabling the continuation for $r < 0$ the vector potential and the
chiral fields. Applying that to superpotential,  we obtain on the negative sheet an anti-DW which
gives just the same contribution to  total mass, but with opposite sign. It means that on the
negative sheet we have an anti-DW, and the DW source of the KN solution may be considered as a
DW-antiDW system similar to the known kink-antikink solutions, called also as
  `breathers', see for example \cite{Lomd}.

  Integration (\ref{MbagKerr}) over
$r\in [-R,R] ,$ shows that the mass-energy of the DW breather is
zero, and the total mass of the KN source will only be determined
by the surface currents and by contribution from the external KN
EM field.

The KN breather regularizes gravitational field of the KN
solution, and we have to specify the regularization of the EM
field now. Taking into account the relation $d\phi_K =d\phi + a(r^2 + a^2)^{-1}dr  $ we note that r- component of the vector-potential
$ er(r^2 + a^2)^{-1} dr $ is regular, and moreover, it is perfect differential and can be dropped.
 Singularity of the vector-potential is places in the equatorial plane $\cos\theta=0$, and outside the source, it takes
 maximal value at the border of the disk, $r= r_e=e^2/2m , \ \cos\theta=0 ,$
where from (\ref{Am}) and (\ref{km}) we obtain
\be A_\m^{(max)}dx^\m = \frac e {r_e}(dt + a \sin ^2 \theta d\phi ).\ee This
border locates very close to the former Kerr singular ring, and the boundary
$r_e=e^2/2m$ appears as the corresponding cut off parameter. This maximal value extends inside the source under the control of the Higgs field according to (\ref{Main}) which is equivalent to Bogomolnyi equations (\ref{mainAH}). As a result, the
vector-potential takes inside the source gradient form, and  the EM strength $F^\mn$ vanishes together with its
contribution to the total mass in (\ref{T3}).
 However, matching of the
internal potential with external EM field turns out to be broken
on the border of source outside the equatorial plane, and there
appear the surface currents $J_\m (r) ,$ penetrating inside the
source  according to (\ref{Main}) or (\ref{JD}).  The
corresponding contribution to the mass density is \be \m_J
(r)=|\cD^{(1)}_t \Phi^1|^2 + |\cD^{(1)}_\phi \Phi^1|^2 = \frac
{|J_t|^2 + |J_\phi|^2} {e^2 \Phi \bar \Phi} . \label{mI}\ee We
note that although we deleted the negative sheet of metric and and
interior of the source is flat, and also all components of the
energy density of the source are regularized,  the
vector-potential $A_\m$ inside the source remains singular and
two-valued because the Kerr coordinate system allows analytic
extension to negative values of the coordinate $r.$

 \textbf{Conclusion. }
We considered supersymmetric source of the KN solution formed by
the N=1  DW-bubble model.  We obtain that this source represents
 indeed the bubble-antibubble system,  analog of the kink-antikink
solution.  The known for planar DW  transformation  to
Bogomolnyi form is generalized to the curved DW-bubble of the Kerr
geometry, where it acquires the space-time dependence demonstrating
its efficiency at full capacity.

 We noted that supersymmetry determines uniquely
 the shape  of the source of the KN solution, and therefore, its stability.
 Any deviation of the domain wall boundary from the surface of ``zero gravity potential'',
 determined by the equation (\ref{HR0}), must break supersymmetry,
 adding gravitating terms to the supersymmetric vacuum states and
breaking the Bogomolnyi bound for the domain wall phase
transition. This mechanism works for any value of the
gravitational constant, and therefore, the Einstein-Maxwell
gravity controls the shape and stability of the source, despite
the known view  on the weakness of the gravitational interaction.

As is known, the Kerr solution describes gravitational field of a
spinning particle, and in particular, since the gyromagnetic ratio
of the KN solution is $g=2 ,$ as that of the Dirac electron,
\cite{Car}, it must correspond to the gravitational and em field
of the electron, and the problem of the source of the KN solution
may not be purely academic, but also it can be related to the problem of the role gravity beyond
Standard Model, shedding the light on the features of the spinning
particles which are required for consistency with gravity. In
particular, it was shown recently ,\cite{BurBag,BurBag2}, that the
source of the KN solution reflects some features of the famous MIT
and SLAC bag models, but the typically used there quartic
potential of the Higgs field conflicts with the external KN
gravity, and the consistency with gravity requires
the supersymmetric scheme of  phase transition. This aspect of the problem of
source of KN solution is beyond the scope of this letter.

 This research is supported by the RFBR grant No. 13-01-00602.

\end{document}